\documentclass[12pt,leqno,letterpaper]{article}

\usepackage{amsmath,amsthm,enumerate,amssymb}
\usepackage[latin1]{inputenc}
\usepackage[T1]{fontenc}
\usepackage[english]{babel}
\usepackage{graphicx}

\newtheorem{theorem}{Theorem}[section]
\newtheorem{proposition}[theorem]{Proposition}

\newtheorem{corollary}[theorem]{Corollary}
\newtheorem{definition}[theorem]{Definition}

\newcommand{\Tr}{{\rm Tr\hskip -0.2em}~}

\newcommand{\pardif}[2]{\frac{\partial#1}{\partial#2}}

\newcommand{\Cov}{\rm Cov}
\newcommand{\Var}{\rm Var}

\begin{document}

\title{Inequalities for quantum skew information}
\author{Koenraad Audenaert, Liang Cai and Frank Hansen}
\date{March 22 2008\\
{\tiny Revised July 20, 2008}}

\maketitle

\begin{abstract}

We study quantum information inequalities and show that the basic
inequality between the quantum variance and the metric adjusted skew
information generates all the multi-operator matrix inequalities or
Robertson type determinant inequalities studied by a number of
authors. We introduce an order relation on the set of functions
representing quantum Fisher information that renders the set into a
lattice with an involution. This order structure generates new
inequalities for the metric adjusted skew informations. In particular,
the Wigner-Yanase skew information is the maximal skew information
with respect to this order structure in the set of Wigner-Yanase-Dyson skew
informations.
\\[1ex]

\end{abstract}

%%%%%%%%%%%%%%%%%%%%%%%%%%%%%%%%%%%%%%%%%%%%%%%%%%%%%%%%%%%%%%%%%%%%%%%%%%%%%%%%%%%%%%%%%%%%%%%%%%%%%%%%%%%%%%%%%%%%%%%%%%

\section{Introduction}

Recently there has been a lot of effort going into the investigation
of various quantum information inequalities. We shall present a
unified view that will cover most of the already known results and
add some more. The notion of metric adjusted skew information was
introduced by the third author \cite{Hansen[1]} as a measure of
quantum information generalizing the Wigner-Yanase-Dyson skew
informations \cite{kn:wigner:1963}. The metric adjusted skew
information is a convex function on the manifold of states and also
satisfies other requirements, suggested by Wigner and Yanase, for a
measure of the information content of a state with respect to a
conserved observable. These requirements include additivity with
respect to the aggregation of isolated subsystems and time
independence for an isolated system. One more requirement, super
additivity, fails spectacularly \cite{kn:hansen:2007} for all the
measures, including the Wigner-Yanase skew information.

We introduce an order relation $ \preceq $ on the set of functions
representing quantum Fisher information that renders the set into a
lattice with a maximal and a minimal element. The order relation induces an order relation on the set of
metric adjusted skew informations that renders also this set into a
lattice with the SLD-information as maximal element. There is no
minimal element.

The Wigner-Yanase-Dyson skew informations given by
\[
I_\rho(p,A)=-\frac{1}{2}\Tr [\rho^p,A][\rho^{1-p},A]\qquad 0<p<1
\]
are shown to be increasing in the parameter $ p $ with respect to this new order
relation in the interval $ (0, 1/2] $ and decreasing in the interval
$ [1/2, 1). $ There is thus maximum in the Wigner-Yanase skew
information for $ p=1/2. $
More elementary, the function $ p\to I_\rho(p,A) $ is increasing in
$ (0,1/2] $ and decreasing in $ [1/2,1) $ for fixed $ \rho $ and $
A. $

\subsection{Basic notations and definitions}

The metric adjusted skew information is defined by setting
\begin{equation}\label{formula: metric adjusted skew information}
    I^c_\rho(A)=\frac{m(c)}{2}\Tr i[\rho, A^*] c(L_\rho,R_\rho)i[\rho, A]
    \end{equation}
    for every positive definite $ n\times n$ density matrix $ \rho $ and every $ n\times n $
    matrix $ A. $ The function $ c $ of two variables is a so called regular Morozova-Chentsov function
    which may be written on the form
\[
c(x,y)=\frac{1}{y f(xy^{-1})}\qquad x,y>0,
\]
where $ f $ is an operator monotone function defined on the positive
half-axis with $ f(1)=1 $ satisfying the functional equation $
f(t)=t f(t^{-1}) $ for $ t>0. $ The regularity condition means that
$ m(c)=f(0)>0. $ The operators $ L_\rho $ and $ R_\rho $ are the
commuting positive definite left- and right- multiplication
operators by $ \rho. $ If we want to emphasize the dependence of the
representing operator monotone function $ f $ we may also denote the
metric adjusted skew information by $ I^f_\rho(A). $ The metric
adjusted skew information is thus proportional to the metric length,
as measured by the quantum Fisher information, of the commutator $
i[\rho, A]. $ In the case of the Wigner-Yanase-Dyson information
this proportionality was already noticed in \cite{kn:petz:1996:1}.
The constant is however important as it can be chosen such that the
metric adjusted skew information coincides with the variance on pure
states.

We have tacitly extended the definition of the metric adjusted skew
information to include the case where $ A $ may not be self-adjoint.
This does not directly make sense in physical applications, but it
turns out to be a useful mathematical tool. The symmetry of $ c $
implies that
\begin{equation}\label{complexification}
I^c_\rho(A+iB)=I^c_\rho(A)+I^c_\rho(B)
\end{equation}
for self-adjoint $ A $ and $ B. $ In particular, the metric adjusted
skew information is convex in $ \rho $ also for non self-adjoint
``observables''.

\subsection{The dynamical uncertainty principle}

The basic quantum information inequality is given by
\begin{equation}\label{basic quantum information inequality}
0\le I^c_\rho(A)\le\Var_\rho(A),
\end{equation}
where the symmetrized variance
\[
\Var_\rho(A)=\frac{1}{2}\Tr\rho(A^*A+AA^*) - |(\Tr\rho A)|^2.
\]
 The inequality was stated and proved in \cite{Hansen[1]} for self-adjoint $ A $ by noting
 that the metric adjusted skew information is a convex function in $ \rho $ while the covariance is concave,
 and that the two measures coincide on pure states. Gibilisco et al. \cite[Proposition 9.2]{gibilisco[2]} subsequently gave another proof.
 The inequality generalizes an earlier result by Luo for the Wigner-Yanase skew information \cite{Luo[5]}. Since the
 symmetrized variance satisfies
 \[
 \Var_\rho (A+iB)=\Var_\rho(A) + \Var_\rho(B)
 \]
 for self-adjoint $ A $ and $ B, $ we immediately obtain the inequality
 (\ref{basic quantum information inequality}) also for non self-adjoint $ A. $

 The above observations effectively reduce the multi-dimensional versions of the quantum information inequalities given by
 a number of authors \cite{gibilisco[4]} to the main inequality (\ref{basic quantum information inequality}).
 Let $ (A_1,\dots, A_k) $ be a tuple of (possibly non self-adjoint) matrices. We then obtain the inequality
 \begin{equation}\label{matrix inequality}
 \Bigl(I_\rho(A_i,A_j)\Bigr)_{i,j=1}^k \le \Bigl(\Cov_\rho(A_i,A_j)\Bigr)_{i,j=1}^k
 \end{equation}
 where
 \[
 I^c_\rho(A,B)=\frac{m(c)}{2}\Tr i[\rho, A^*] c(L_\rho,R_\rho)i[\rho, B]
 \]
 and
\[
\Cov_\rho(A,B)=\frac{1}{2}\Tr\rho(A^*B+BA^*) - (\Tr\rho A^*)(\Tr\rho
B).
\]
Indeed, we only have to notice the identity
\[
I_\rho^{c}(\xi_1 A_1+\cdots+\xi_k A_k)=
\left(\Bigl(I_\rho(A_i,A_j)\Bigr)_{i,j=1}^k\xi\mid \xi\right),
\qquad\xi=\begin{pmatrix}
\xi_1\\
\vdots\\
x_k
\end{pmatrix},
\]
for arbitrary complex numbers $ \xi_1,\dots,\xi_k, $ and the similar
identity for the covariance matrix. This reduces the matrix version
(\ref{matrix inequality}) to the basic quantum information
inequality (\ref{basic quantum information inequality}). In
particular, we obtain the determinant inequality
\begin{equation}\label{determinant inequality}
 0\le\det\Bigl(I_\rho(A_i,A_j)\Bigr)_{i,j=1}^k \le \det\Bigl(\Cov_\rho(A_i,A_j)\Bigr)_{i,j=1}^k
\end{equation}
by the well known formula $ \det A=\exp\Tr\log A, $ since the
logarithm is operator monotone\footnote{In fact, the trace function $ A\to\Tr f(A) $ is increasing for any increasing function $ f. $}.
This version has been coined the ``dynamic uncertainty principle''.

\section{Inequalities for quantum skew information}

We introduce various methods to compare measures of quantum
information for one observable. The generalization to several
observables is then obtained as explained in the introduction.

The metric adjusted skew information may be written
\cite[Proposition 3.4]{Hansen[1]} on the form
\begin{equation}\label{skew information in terms of d}
    I^c_\rho(A)=\Tr\rho A^2-\frac{m(c)}{2}\Tr A\, d_c(L_\rho,R_\rho)A,
\end{equation}
where
\[
d_c(x,y) =\frac{x+y}{m(c)}-(x-y)^2 c(x,y)\qquad x,y>0
\]
is operator concave. Since $ d_c $ is homogeneous of degree one we
may write
\[
\frac{m(c)}{2}\, d_c(x,y)=y \tilde f(xy^{-1}),
\]
where the function
\[
\tilde f(t)=\frac{f(0)}{2}\, d_c(t,1)=\frac{t+1}{2} -
(t-1)^2\frac{f(0)}{2f(t)}.
\]
Since $ \tilde f $ is operator concave and defined on the positive
half-axis it is also operator monotone \cite{kn:hansen:1982}. With
this notation the metric adjusted skew information takes the form
\begin{equation}\label{skew information in terms of tilde f}
    I^c_\rho(A)=\Tr\rho A^2-\Tr A\, R_\rho \tilde f(L_\rho R_\rho^{-1})A
\end{equation}
as it is studied in \cite{Petz[1]}.

We begin by providing a more detailed classification of the
representing functions for quantum fisher information.

\begin{theorem}\label{theorem: canonical representation of f}
    Let $ f\colon\mathbf R_+\to\mathbf R_+ $ be a function satisfying
    \begin{enumerate}[(i)]
    \item $ f $ is operator monotone,
    \item $ f(t)=tf(t^{-1}) $ for all $ t>0, $
    \item $ f(1)=1. $
    \end{enumerate}
    Then $ f $ admits a canonical representation
    \begin{gather}\label{canonical representation of f}
    f(t)=\frac{1+t}{2}\exp\int_0^1\frac{(\lambda^2-1)(1-t)^2}{(\lambda+t)(1+\lambda t)(1+\lambda)^2}\,h(\lambda)\,d\lambda
    \end{gather}
    where the weight function $ h:[0,1]\to[0,1] $ is measurable.
    The equivalence class containing $ h $ is uniquely determined by $ f. $ Any function
    on the given form maps the positive half-axis into itself and satisfy the conditions
    $ (i), $ $ (ii) $ and $ (iii). $
    \end{theorem}

\begin{proof}
The third author \cite[Theorem 1]{kn:hansen:2006:2} gave an
exponential representation of the functions $ f\colon\mathbf
R_+\to\mathbf R_+ $ satisfying $ (i) $ and $ (ii) $ of the form
\begin{equation}
f(t)=e^\beta\frac{1+t}{\sqrt{2}}\exp\int_0^1\frac{\lambda^2-1}{\lambda^2+1}\cdot
\frac{1+t^2}{(\lambda+t)(1+\lambda t)}\,h(\lambda)\,d\lambda
\end{equation}
where $ h:[0,1]\to[0,1] $ is measurable and $
\exp\beta=f(i)\exp(-i\pi/4). $ Setting
\[
f(1)=e^\beta\frac{2}{\sqrt{2}}\exp\int_0^1\frac{\lambda^2-1}{\lambda^2+1}\cdot
\frac{2}{(\lambda+1)^2}\,h(\lambda)\,d\lambda=1
\]
and solving for $ \exp\beta $ we obtain
\[
f(t)=\frac{1+t}{2}\exp\int_0^1\frac{\lambda^2-1}{\lambda^2+1}
\left(\frac{1+t^2}{(\lambda+t)(1+\lambda t)} -
\frac{2}{(\lambda+1)^2}\right) h(\lambda)\,d\lambda
\]
which reduces to the expression in the theorem. The uniqueness of $
h $ (up to equivalence) and the sufficiency of the condition follows
from the reference.
\end{proof}

Following several authors we denote by $ \mathcal F_{\text{op}} $
the set of functions characterized in Theorem~\ref{theorem:
canonical representation of f}. A function $ f\in\mathcal
F_{\text{op}} $ is said to be regular if its extension to the closed
positive half line satisfies $ f(0)> 0. $ If $ f\in  \mathcal
F_{\text{op}} $ is regular we set
\begin{equation}
\check f(t)=\frac{f(0)}{f(t)}\quad\text{and}\quad \check
c(x,y)=y^{-1} \check f(xy^{-1})
\end{equation}
and may write
\[
I_\rho^f(A)=\frac{1}{2}\Tr i[\rho, A^*] \check
c(L_\rho,R_\rho)i[\rho, A].
\]
Notice that $ \check c $ is a symmetric function in two positive
variables, and that the metric adjusted skew information $
I_\rho^f(A) $ is increasing in the transform $ \check f, $ cf. also
\cite{gibilisco[4],gibilisco[5],gibilisco[3]}.

The representing function $ f_p\in\mathcal F_{\text{op}} $ of the
Wigner-Yanase-Dyson skew information $ I_\rho(p, A) $ with parameter $ p $ is given by
\[
f_p(t)=p(1-p)\cdot\frac{(t-1)^2}{(t^p-1)(t^{1-p}-1)}\qquad 0<p<1.
\]
We observe that the transform
\[
\check f_p(t)=\frac{(t^p-1)(t^{1-p}-1)}{(t-1)^2}=\frac{t+1-(t^p+t^{1-p})}{(t-1)^2}
\]
is increasing in $ p\in(0,1/2] $ and decreasing in $ p\in[1/2,1). $ It follows that
the Wigner-Yanase-Dyson skew information $ I_\rho(p,A), $ for fixed $ \rho $ and $ A, $
is an increasing function of $ p $ in the interval $ (0,1/2] $ and a decreasing function
of $ p $ in the interval $ [1/2,1) $ with maximum in the Wigner-Yanase skew information.

\begin{proposition}\label{representation of tilde f}
The transform $ \check f $ of a regular function $ f\in\mathcal
F_{\text{op}} $ has a canonical representation
\begin{equation}\label{f(0)/f(t)}
\check f(t)=\frac{f(0)}{f(t)}=\frac{1}{(1+t)}\exp\int_0^1
\frac{t(\lambda^2-1)}{\lambda(\lambda+t)(1+\lambda t)}\,
h(\lambda)\,d\lambda,
\end{equation}
where the weight function $ h:[0,1]\to[0,1] $ is measurable and
\[
\int_0^1 \frac{h(\lambda)}{\lambda}\,d\lambda<\infty.
\]
The equivalence class containing $ h $ is uniquely determined by $
f. $ Any function on the given form is the transform $ \check f $ of
a regular function $ f $ in $ \mathcal F_{\text{op}}. $
\end{proposition}

\begin{proof}
The integrability condition for a weight function $ h $ of a regular
function $ f\in\mathcal F_{\text{op}} $ follows from
Theorem~\ref{theorem: canonical representation of f}. Furthermore,
\[
f(0)=\frac{1}{2}\exp\int_0^1\frac{(\lambda^2-1)}{\lambda
(1+\lambda)^2}\,h(\lambda)\,d\lambda
\]
and thus
\[
\frac{f(0)}{f(t)}=\frac{1}{(1+t)}\exp\int_0^1
\left(\frac{(\lambda^2-1)}{\lambda (1+\lambda)^2}-
\frac{(\lambda^2-1)(1-t)^2}{(\lambda+t)(1+\lambda
t)(1+\lambda)^2}\right) h(\lambda)\,d\lambda
\]
which reduces to the expression in the proposition.
\end{proof}

Since the integrand in (\ref{f(0)/f(t)}) is non-positive, we realize
that $ \check f $ and thus the metric adjusted skew information is
decreasing in the weight function $ h. $

\begin{definition}\label{definition: preceq}
Let $ f,g\in \mathcal F_{\text{op}} $ and set
\[
\varphi(t)=\frac{t+1}{2}\frac{f(t)}{g(t)}\qquad t>0.
\]
We write $ f\preceq g $ if the function $ \varphi\in \mathcal
F_{\text{op}}. $
\end{definition}

The function $ \varphi $ obviously satisfy $ \varphi(1)=1 $ and $
\varphi(t)=t\varphi(t^{-1}) $ for $ t>0. $ The condition in the
definition is thus equivalent to operator monotonicity of $ \varphi.
$ Setting
\[
f_{\min}(t)=\frac{2t}{t+1}\quad\text{and}\quad
f_{\max}(t)=\frac{1+t}{2}\qquad t>0,
\]
it is known that $ f_{\min}\le f\le f_{\max} $ for every function $
f\in\mathcal F_{\text{op}}. $ But we realize that also
\begin{equation}
f_{\min}\preceq f\preceq f_{\max}\qquad\text{for every}\quad
f\in\mathcal F_{\text{op}}.
\end{equation}
This is so since $ f_{\min}\preceq f $ is reduced to operator
monotonicity of the function $ t\to t f(t)^{-1}, $ while $ f\preceq
f_{\max} $ is reduced to operator monotonicity of $ f $ itself.
Since as noted $ \varphi\le f_{\max} $ for $ \varphi\in \mathcal
F_{\text{op}} $ we realize that
\begin{equation}
f\preceq g\quad\Rightarrow\quad f\le g\qquad\text{for}\quad f,g\in
\mathcal F_{\text{op}}.
\end{equation}
In particular, $ f\preceq g $ and $ g\preceq f $ implies $ f=g. $

\begin{theorem}\label{theorem: partial order relation on F_op}
The relation $ \preceq $ is a partial order relation on $ \mathcal
F_{\text{op}} $ rendering $ (\mathcal F_{\text{op}}, \preceq) $ into
a lattice with $ f_{\min} $ as the minimal element and $ f_{\max} $
as the maximal element. Furthermore, if $ f,g\in \mathcal
F_{\text{op}} $ then
\[
f\preceq g\quad \Leftrightarrow\quad h_f \geq h_g\qquad\text{a.e.},
\]
where $ h_f $ and $ h_g $ respectively are representing functions of
$ f $ and $ g $ as in Theorem~\ref{theorem: canonical representation
of f}.
\end{theorem}

\begin{proof} The function $ \varphi $ in Definition~\ref{definition: preceq} has the representation
\[
 \varphi(t)=\frac{1+t}{2}\exp\int_0^1
 \frac{(\lambda^2-1)(1-t)^2}{(\lambda+t)(1+\lambda t)(1+\lambda)^2}\,\bigl(h_f(\lambda)-h_g(\lambda)\bigr)\,d\lambda
\]
and is therefore operator monotone, according to
Theorem~\ref{theorem: canonical representation of f}, if and only if
$ 0\le h_f(\lambda)-h_g(\lambda)\le 1 $ for almost all $ \lambda\in
[0,1]. $ Therefore, $ f\preceq g $ if and only if $ h_f\geq h_g $
almost everywhere.

It follows that $ \preceq $ is an order relation. Indeed, if $
f_1\preceq f_2 $ and $ f_2\preceq f_3 $ then, for representing
functions, $ h_{f_1}\geq h_{f_2} $ and $ h_{f_2}\geq h_{f_3} $
almost everywhere. Consequently, $ h_{f_1}\geq h_{f_3} $ almost
everywhere and thus $ f_1\preceq f_3. $

For arbitrary $ f,g\in \mathcal F_{\text{op}} $ with representing
functions $ h_f $ and $ h_g $ we define $ f\wedge g $ as the
function in $ \mathcal F_{\text{op}} $ with representing function $
\max\{h_f, h_g\}. $ Similarly, we define $ f\vee g $ as the function
in $ \mathcal F_{\text{op}} $ with representing function $
\min\{h_f, h_g\}. $ It follows that
\[
f\wedge g \preceq f \preceq f\vee g\quad\text{and}\quad f\wedge g
\preceq g \preceq f\vee g.
\]
If furthermore $ \psi\preceq f $ and $ \psi\preceq g $ for a $
\psi\in\mathcal F_{\text{op}} $ it follows that $ \psi\preceq
f\wedge g. $ If similarly $ f\preceq\psi $ and $ g\preceq\psi $ it
follows that $ f\vee g\preceq\psi. $ Therefore $ (\mathcal
F_{\text{op}}, \preceq) $ is a lattice.
\end{proof}

Notice that for $ f,g\in \mathcal F_{\text{op}} $ we have the
relations
\[
f\wedge g \le\min\{f,g\}\le f \le\max\{f,g\} \le f\vee g,
\]
with a similar statement for $ g. $

We next show that the lattice $ \mathcal F_{\text{op}} $ is equipped
with a natural involution.

\begin{definition}
We define for $ f\in\mathcal F_{\text{op}} $ the function
\begin{equation}
f^\sharp(t)=\frac{t}{f(t)}\qquad t>0.
\end{equation}
\end{definition}
It follows from the general theory of operator monotone functions
\cite[2.6. Corollary]{kn:hansen:1982} that $ f^\sharp $ is operator
monotone, and since obviously $ f^\sharp(1)=1 $ and $ f^\sharp(t)=t
f^\sharp(t^{-1}) $ we realize that also $ f^\sharp\in\mathcal
F_{\text{op}}. $

It also follows that $ f^{\sharp\sharp}=f $ and that $ f(t)=t^{1/2}
$ is the unique fixpoint of the involution $ f\to f^\sharp $ in $
\mathcal F_{\text{op}}. $ Furthermore, $ f\preceq g $ implies $
g^\sharp\preceq f^\sharp $ for functions $ f,g\in\mathcal
F_{\text{op}}. $ We notice that $ f^\sharp $ is not regular for a
regular $ f\in \mathcal F_{\text{op}}. $ All of these assertions may
also be verified by calculating the representing weight functions.

\begin{proposition}
Let $ f\in\mathcal F_{\text{op}} $ with representing weight function
$ h $ as in Theorem~\ref{theorem: canonical representation of f}.
Then $ f^\sharp\in\mathcal F_{\text{op}} $ with representing weight
function $ 1-h. $
\end{proposition}

\begin{proof}
We first write
\[
f^\sharp
(t)=\frac{t}{f(t)}=\frac{2t}{t+1}\exp\left[-\int_0^1\frac{(\lambda^2-1)(1-t)^2}{(\lambda+t)(1+\lambda
t)(1+\lambda)^2}\,h(\lambda)\,d\lambda\right]
\]
and since $ f_{\min}(t)=2t(t+1)^{-1} $ has $ 1 $ as representing
function, we obtain
\[
f^\sharp
(t)=\frac{t+1}{2}\exp\int_0^1\frac{(\lambda^2-1)(1-t)^2}{(\lambda+t)(1+\lambda
t)(1+\lambda)^2}\,(1-h(\lambda))\,d\lambda
\]
and the assertion is proved.
\end{proof}

We obtain from Proposition~\ref{representation of tilde f} and the
preceding remarks:

\begin{corollary}\label{corollary: order relation on metric adjusted skew informations}
The restriction of the order relation $ \preceq $ to the regular
part of $ \mathcal F_{\text{op}} $ induces an order relation on the
set of metric adjusted skew informations.
\end{corollary}

\subsection{Optimality of the Wigner-Yanase information}

The order relation on the set of metric adjusted skew informations introduced in Corollary~\ref{corollary: order relation on metric adjusted skew informations}
is tractable in the sense that we only have to study and compare the representing functions of the associated quantum Fisher informations as they are given in
Theorem~\ref{theorem: canonical representation of f}.

The function $ f_p\in \mathcal F_{\text{op}} $
representing the Wigner-Yanase-Dyson skew information with parameter
$ p\in(0,1) $ has weight function
\[
h_p(\lambda)=\frac{1}{\pi}\arctan\frac{(\lambda^p +
\lambda^{1-p})\sin p\pi}{1-\lambda-(\lambda^p - \lambda^{1-p})\cos
p\pi}\qquad 0<\lambda<1,
\]
according to the representation given in Theorem~\ref{theorem: canonical representation of f},
cf. \cite[Theorem 2.7]{Hansen[1]}.

\begin{theorem}\label{theorem: optimality of WY}
The functions $ h_p(\lambda) $ are decreasing in $ p\in(0,1/2] $ for any $ \lambda $ with $ 0<\lambda<1. $
\end{theorem}

\begin{proof}

We consider a fixed $ \lambda\in(0,1), $ set $ z_0=-\lambda+i\varepsilon $ for a small $ \varepsilon>0, $
and obtain
\[
h_p(\lambda)=-\frac{1}{\pi}\lim_{\varepsilon\to 0} f_\varepsilon(p),
\]
where
\[
\begin{array}{rl}
f_\varepsilon(p) &=\arg((1-z_0^p)(1-z_0^{1-p})) \\[1ex]
&= \arg(1-z_0^p)+\arg(1-z_0^{1-p}).
\end{array}
\]
Obviously, we have $f_\varepsilon(p)=f_\varepsilon(1-p)$.
Thus, to show that $f_\varepsilon(p)$ is increasing over the left half $[0,1/2]$ of the domain,
it suffices to show that the first term $\arg(1-z_0^p)$ is concave in $p$ over the interval $[0,1]$.

We can show that the function $g(p)=\arg(1-z^p)$ is concave over $[0,1]$ for any $z$ in the ``lune'' $ \mathcal I=\{z: |z|<1,\Im z>0\}$,
including $z_0$ for sufficiently small $ \varepsilon>0. $
Since $\arg=\Im\log$, the second derivative $g''(p)$ is given by
\[
g''(p)= \Im\frac{-z^p(\log z)^2}{(1-z^p)^2}=\frac{1}{p^2} \Im\frac{-z^p(\log z^p)^2}{(1-z^p)^2}
\]
and we have to show that this is non-positive for $z\in \mathcal I $ and $0\le p\le 1$.
In fact, it is enough to do this for $p=1$ only, since for $z\in\mathcal I$ and $0\le p\le 1$,
then $z^p\in \mathcal I$ too.

The imaginary part of a complex number is non-positive if and only
its complex argument is between $ \pi $ and $2\pi$. Thus we need to show
\[
q(z)=\arg\frac{-z(\log z)^2}{(1-z)^2} \in[\pi,2\pi]\qquad \forall z\in\mathcal I.
\]
Putting $z=r\exp(i\theta)$, with $0< r\le 1$ and $ 0\le \theta\le \pi $,
\[
\begin{array}{rl}
q(z) &= \arg(-z) +2\arg\log z-2\arg(1-z) \\[1ex]
&\displaystyle =\pi+ \theta + 2\arctan\frac{\theta}{\log r} +2\arctan\frac{r\sin\theta}{1-r\cos\theta}\,.
\end{array}
\]
For $\theta=0$ ($z$ real) $ q(z) $ is obviously $\pi$.
For $\theta=\pi, $
\[
q(z)=2\pi + 2\arctan\frac{\pi}{\log r}<2\pi
\]
for $0<r< 1$.
We can now show that for fixed $r$, $q(z)$ increases with $\theta$, which implies that
$q$ is indeed between $ \pi $ and $2\pi$ for $z\in \mathcal I$.
The first derivative is
\[
\pardif{q}{\theta} = 2\frac{\log r}{(\log r)^2+\theta^2}+\frac{1-r^2}{1+r^2-2r\cos\theta}\,.
\]
Because of the inequality $\cos\theta\ge 1-\theta^2/2$, we obtain a lower bound
\[
\begin{array}{rl}
\displaystyle\pardif{q}{\theta}&\displaystyle\ge\frac{1-r^2}{(1-r)^2 + r\theta^2} +2\frac{\log r}{(\log r)^2+\theta^2}\\[3ex]
&=\displaystyle\frac{\phi(r)\log r +\theta^2 \psi(r)}{((1-r)^2+r\theta^2)((\log r)^2+\theta^2)}
\end{array}
\]
where
\[
\begin{array}{rl}
\phi(r)&=(1-r^2)\log r + 2(1-r)^2,\\[1ex]
\psi(r)&=1-r^2+2r\log r.
\end{array}
\]
The first derivative $ \psi'(r)=2(1-r+\log r) $ and the second derivative $ \psi''(r)=2(1/r-1) $ is
positive. Therefore $ \psi' $ is increasing and since $ \psi'(1)=0, $ we derive that $ \psi' $ is negative and thus $ \psi $ is decreasing. But since $ \psi(1)=0, $ we derive that $ \psi(r)\ge 0 $ for $ 0< r< 1. $ We furthermore calculate
\[
\begin{array}{rl}
\phi'(r)&=\displaystyle-2r\log r +\frac{1}{r}+3r-4,\\[1ex]
\phi''(r)&=\displaystyle\frac{1}{r^2}(r^2-2r^2\log r-1).
\end{array}
\]
The parenthesis in the expression for $ \phi''(r) $ has derivative $ -4r\log r\ge 0 $ and value zero in $ r=1. $
It is thus non-positive, so $ \phi $ is concave and $ \phi' $ therefore
decreasing. Since $ \phi'(1)=0 $ we derive that $ \phi'\ge 0, $ so $ \phi $ is increasing and since $ \phi(1)=0, $
we conclude that $ \phi $ is non-positive. We have consequently proved
\[
\pardif{q}{\theta}\ge0
\]
for $ 0<\theta<\pi $ and $ 0<r<1, $ and the proof is complete.
\end{proof}
Theorem~\ref{theorem: optimality of WY} is illustrated by the graph in Figure 1.
\begin{figure}
\centering
\includegraphics[scale=1]{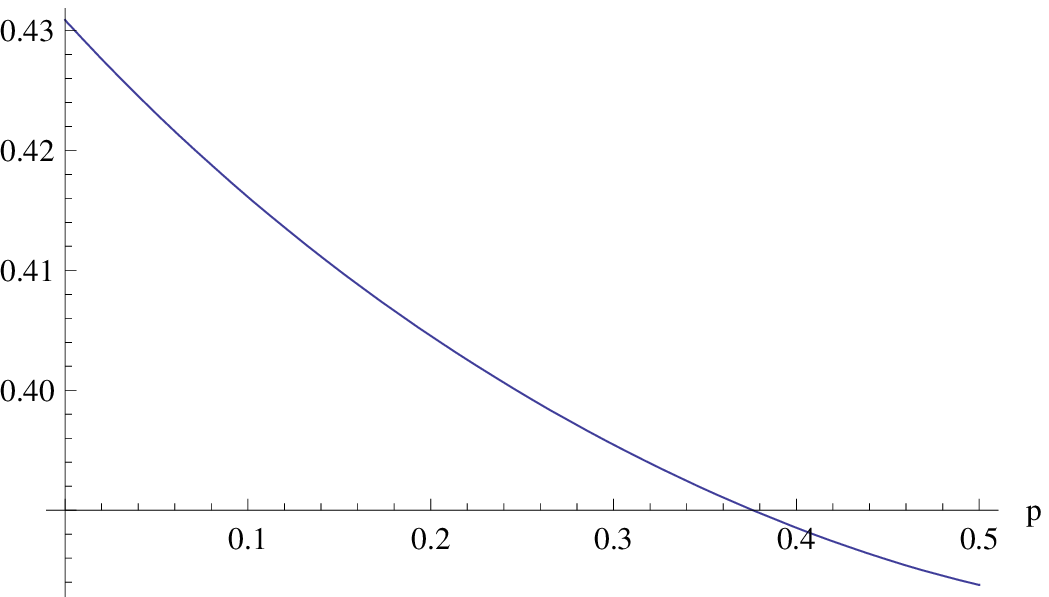}
\caption{$ p\to h_p(\lambda) $ for $ \lambda=1/2$}
\end{figure}
The graphs for different $ \lambda \in(0,1) $ look very much alike
except for a change in units. An equivalent formulation of the
result is that the function
\[
\varphi(t)=\frac{t+1}{2}\frac{f_p(t)}{f_q(t)}=\frac{p(1-p)}{2q(1-q)}\cdot
\frac{(t+1)(t^q-1)(t^{1-q}-1)}{(t^p-1)(t^{1-p}-1)}\qquad t>0.
\]
is operator monotone (and thus belongs to $ \mathcal F_{\text{op}}) $
for $ 0< p\le q\le 1/2. $

We have thus proved that $ f_p\preceq f_q $ for $
0<p\le q\le 1/2 $ and $ f_p \succeq f_q $ for $ 1/2\le p\le q<1. $
The Wigner-Yanase skew information is therefore the
maximal element among the Wigner-Yanase-Dyson informations with
respect to the order relation inherited from $ (\mathcal
F_{\text{op}},\preceq). $

 Another example is the variant bridge considered in
\cite{Hansen[1]} with representing functions $ f_p\in\mathcal
F_{\text{op}} $ with weight functions according to the representation in
Theorem~\ref{theorem: canonical representation of f} given by
 \[
 h_p(\lambda)=\left\{\begin{array}{lrl}
                        0,\quad &\lambda &<1-p\\[1ex]
                        p, &\lambda&\ge 1-p
                        \end{array}\right.\qquad 0\le p\le 1.
 \]
This family of weight functions is decreasing in the parameter $
p\in[0,1]. $ The corresponding quantum Fisher informations are therefore
increasing in the parameter $ p $ and connects the SLD-metric for $
p=0 $ with the  Bures metric for $ p=1, $ and they are regular for $
p<1. $ The corresponding metric adjusted skew informations, defined
for $ p\in[0,1), $ are increasing with respect to the order relation
$ \preceq. $

%%%%%%%%%%%%%%%%%%%%%%%%%%%%%%%%%%%%%%%%%%%%%%%%%%%%%%%%%%%%%%%%%%%%%%%%%%%%%%%%

%%%%%%%%%%%%%%%%%%%%%%%%%%%%%%%%%%%%%%%%%%%%%%%%%%%%%%%%%%%%%%%%%%%%%%%%%%%%%%%%%%%%%%%%%%%%%%%%%%%%%%%%%%%%%%%%%%%%%%%%%%
{\small

%      \bibliographystyle{plain}
%    \bibliography{mathharv}

\vfill

    \noindent Koenraad Audenaert: Mathematics department of Royal Holloway, University of London Egham Hill, EGHAM TW20 0EX, UK.\\[1ex]
      \noindent Liang Cai: Academy of Mathematics and Systems
       Science, Chinese Academy of Sciences, Beijing 100080, China.\\[1ex]
      \noindent Frank Hansen: Department of Economics, University
       of Copenhagen, Studiestraede 6, DK-1455 Copenhagen K, Denmark.}

\end{document}